\documentclass[12pt]{article}
\usepackage{graphicx}
\usepackage{lineno,xcolor}
\usepackage{cite}
\usepackage{xspace}
\newcommand{\Xicc}{$\Xi_{cc}^{+}$}
\newcommand{\Lc}{$\Lambda_{c}^{+}$}
\newcommand{\tev}{\ensuremath{\mathrm{\,Te\kern -0.1em V}}\xspace}
\newcommand{\mevcc}{\ensuremath{{\mathrm{\,Me\kern -0.1em V\!/}c^2}}\xspace}
\newcommand{\mev}{\ensuremath{\mathrm{\,Me\kern -0.1em V}}\xspace}
\newcommand{\mevc}{\ensuremath{{\mathrm{\,Me\kern -0.1em V\!/}c}}\xspace}
\newcommand{\dm}{\ensuremath{\delta m}}

\def\pbnr{}
\def\speaker{Stephen Ogilvy}
\def\onbehalfof{the LHCb collaboration}
\def\title{Studies of charmed baryons at LHCb}
\def\affiliation{School of Physics and Astronomy\\
The University of Glasgow, Glasgow, UK}
\def\support{The workshop was supported by the University of Manchester, IPPP, STFC, and IOP}

\newcommand\pubnumber{\pbnr}
\newcommand\pubdate{\today}
\def\Title#1{\begin{center} {\Large #1 } \end{center}}
\def\Author#1{\begin{center}{ \sc #1} \end{center}}

\newcommand{\OnBehalf}[1]{\sbox0{#1}\ifdim\wd0=0pt
        {}
	\else
	{\\on behalf of #1}
	\fi}
\newcommand{\SupportedBy}[1]{\sbox0{#1}\ifdim\wd0=0pt
        {}
	\else
	{\footnote{#1}}
	\fi}
\def\Address#1{\begin{center}{ \it #1} \end{center}}

\newcommand\pubblock{\includegraphics[width=5cm]{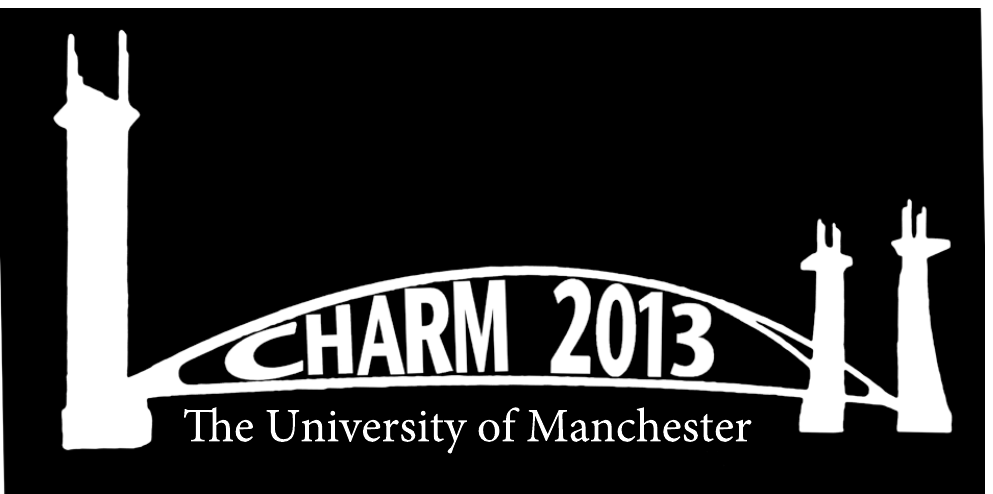}\hfill{\begin{tabular}{l} \pubnumber\\
         \pubdate  \end{tabular}}}
\newenvironment{Abstract}{\begin{quotation}  }{\end{quotation}}
\newenvironment{Presented}{\begin{quotation} \begin{center} 
             PRESENTED AT\end{center}\bigskip 
      \begin{center}\begin{large}}{\end{large}\end{center} \end{quotation}}
\def\Acknowledgements{\bigskip  \bigskip \begin{center} \begin{large}
             \bf ACKNOWLEDGEMENTS \end{large}\end{center}}
\def\venue{The 6$^{\mathrm{th}}$ International Workshop on Charm Physics\\
(CHARM 2013)\\
Manchester, UK,  31 August -- 4 September, 2013}

\def\beq{\begin{equation}}
\def\eeq#1{\label{#1}\end{equation}}
\def\eeqn{\end{equation}}

\def\beqa{\begin{eqnarray}}
\def\eeqa#1{\label{#1}\end{eqnarray}}
\def\eeqan{\end{eqnarray}}

\let\bar=\overbar

\def\Dslash{\not{\hbox{\kern-4pt $D$}}}
\def\dslash{\not{\hbox{\kern-2pt $\del$}}}

\def\msb{{\bar{\ssstyle M \kern -1pt S}}}

\begin{document}
\begin{titlepage}
\pubblock

\vfill
\Title{\title}
\vfill
\Author{\speaker\SupportedBy{\support}\OnBehalf{\onbehalfof}}
\Address{\affiliation}
\vfill

\begin{Abstract}
We report a search for the doubly charmed baryon $\Xi_{cc}^{+}$ through the decay $\Xi_{cc}^{+} \to \Lambda_{c}^{+} K^{-} \pi^{+}$, using a data sample corresponding to an integrated luminosity of $0.65~\mathrm{pb^{-1}}$ of $pp$ collisions at $\mathrm{\sqrt{s} = 7\tev}$. In the mass range 3300--3800\mevcc no significant signal is observed. Upper limits at $95\%$ confidence level are set on $R$, the ratio of the production cross section of the \Xicc\ times the relevant branching fraction over the $\Lambda_{c}^{+}$ cross section, as a function of the \Xicc\ mass and lifetime. The largest upper limits on $R$ over the investigated mass range are $R<1.5\times10^{-2}$ for a lifetime of $100~\mathrm{fs}$ and $R<3.9\times10^{-4}$ for a lifetime of $400~\mathrm{fs}$.
\end{Abstract}


\vfill
\begin{Presented}
\venue
\end{Presented}
\vfill
\end{titlepage}
\def\thefootnote{\fnsymbol{footnote}}
\setcounter{footnote}{0}
%

\section{Introduction}
For the four lightest quarks predicted in the constituent quark model, the baryonic states are predicted to form $SU(4)$ multiplets. For the ground states with $C=2$, a $\Xi_{cc}$ isodoublet $(ccu,ccd)$ and an $\Omega_{cc}$ isosinglet are expected. There are numerous predictions of the properties of these states, with the majority yielding masses in the range 3500--3700\mevcc and a lifetime in the range 100--250$~\mathrm{fs}$ \cite{Roberts:2007ni, He:2004px, Wang:2010hs, Chang:2006eu, Valcarce:2008dr, Chang:2007xa, Ebert:2002ig, Guberina:1999mx}. The only observed signals for any of these states are those reported by the SELEX experiment for the \Xicc\ in its decays to $\Lambda_{c}^{+} K^{-} \pi^{+}$ and $p D^{+} K^{-}$ \cite{Mattson:2002vu,Ocherashvili:2004hi}. The reported state had a mass measured to be $\mathrm{3519\pm2~MeV/c^{2}}$ and a lifetime consistent with zero, and less than $\mathrm{33fs}$ at the $90\%$ confidence level. Subsequent searches at the BELLE \cite{Chistov:2006zj} and BaBar \cite{Aubert:2006qw} experiments have not observed any evidence for doubly charmed baryon production. In these proceedings, we report the results of a search for the \Xicc\ baryon at LHCb \cite{Aaij:2013voa}.

\section{Analysis method}
For comparison with subsequent searches in hadronic environments we measure the \Xicc\ production relative to that of the $\Lambda_{c}^{+}$:
\begin{equation}
R \equiv \frac{\sigma(\Xi_{cc}^{+})\mathcal{B}(\Xi_{cc}^{+} \to \Lambda_{c}^{+}K^{-}\pi^{+})}{\sigma(\Lambda_{c}^{+})} = \frac{N_{\mathrm{sig}}}{N_{\mathrm{norm}}}\frac{\epsilon_{\mathrm{norm}}}{\epsilon_{\mathrm{sig}}}
\end{equation}
where $\sigma$ and $\mathcal{B}$ represent cross sections and branching fractions, respectively, $N_{\mathrm{sig}}$ and $N_{\mathrm{norm}}$ are the extracted yields of the \Xicc\ signal and the control \Lc, and $\epsilon_{\mathrm{sig}}$ and $\epsilon_{\mathrm{norm}}$ are the efficiencies of those modes. A reasonable expectation is that \\ $\mathcal{B}(\Xi_{cc}^{+} \to \Lambda_{c}^{+}K^{-}\pi^{+}) \approx \mathcal{B}(\Lambda_{c}^{+} \to p^{+}K^{-}\pi^{+}) \approx 5\%$. The LHCb \Lc\ cross-section at $\mathrm{\sqrt{s}} = 7$ \tev\ has been measured to be $230 \pm 77 ~\mathrm{\mu b}$ \cite{Aaij:2013mga}. Phenomenological estimates of the \Xicc\ production cross section in a $pp$ environment at $\mathrm{\sqrt{s}} =$ 14 \tev\ range between 60--1800 $\mathrm{nb}$ \cite{Chang:2006eu}, and at $\mathrm{\sqrt{s} = 7}$ \tev\ this is expected to be approximately halved. Therefore at LHCb $R$ is expected to be of order $10^{-5} - 10^{-4}$.

To account for the a priori unknown \Xicc\ mass and lifetime we search for the \Xicc\ in a wide mass range ($\mathrm{3300-3800~MeV/c^{2}}$) and calculate  efficiencies for a variety of lifetime hypotheses. For each candidate the mass difference is calculated as
\begin{equation}
\delta m \equiv m([pK^{-}\pi^{-}]_{\Lambda_{c}^{+}}K^{-}\pi^{+}) - m([pK^{-}\pi^{-}]_{\Lambda_{c}^{+}}) - m(K^{-}) - m(\pi^{+}) 
\end{equation}
where $m([pK^{-}\pi^{-}]_{\Lambda_{c}^{+}}K^{-}\pi^{+})$ is the measured invariant mass of the reconstructed \Xicc\ candidate, $m([pK^{-}\pi^{-}]_{\Lambda_{c}^{+}})$ is the measured mass of the reconstructed \Lc\ candidate and $m(K^{-})$ and $m(\pi^{+})$ are respectively the charged kaon and pion world-averaged masses. This \Xicc\ mass window corresponds to a $\delta m$ signal window of $380 < \delta m < 880$ \mev. 

Our analysis is carried out using a data sample corresponding to an integrated luminosity of $0.65~\mathrm{pb^{-1}}$ of $pp$ collisions at $\mathrm{\sqrt{s}} = 7$ \tev, from the data gathered at LHCb during 2011. The analysis procedure was fixed before the data in the signal region was examined. Limits are on $R$ are given as a function of both the \Xicc\ mass and lifetime.

\section{Candidate selection}
The selection procedure to trigger, reconstruct and select candidates must retain signal candidates and suppress three main sources of background. These backgrounds are combinations of unrelated tracks, mis-reconstructed heavy-flavour decays, and combinations of a real \Lc\ with unrelated tracks. The first two lead to smooth distributions in both $m([pK^{-}\pi^{-}]_{\Lambda_{c}^{+}})$ and  $\delta m$, while the third background only peaks in $m([pK^{-}\pi^{-}]_{\Lambda_{c}^{+}})$ and is smooth in $\delta m$.

The selection in the software and hardware triggers for the signal and normalisation mode ($\Lambda_{c}^{+} \to pK^{-}\pi^{+}$) is identical to reduce systematic uncertainties. A candidate must fulfil the criteria that one of the three \Lc\ daughter tracks must be associated with a calorimeter cluster with a measured transverse energy greater than 3500 \mev\ to fire the hardware trigger. One of the \Lc\ daughter tracks must then be selected by an inclusive selection algorithm in the software trigger, which requires the track possesses a transverse momentum greater than 1700 \mevc and $\chi^{2}_{\mathrm{IP}} > 16$ with respect to any primary vertex, where $\chi^{2}_{\mathrm{IP}}$ is the increase to the associated primary vertex's reconstructed $\chi^{2}$ when the track is included in the primary vertex fit.

The \Lc\ candidate must then be reconstructed by a dedicated $\Lambda_{c}^{+} \to p K^{-} \pi^{+}$ selection algorithm which makes a variety of kinematic and geometric requirements. The candidate must be displaced from the primary vertex, the reconstructed $\Lambda_{c}^{+} ~ p_{T} > 500$ \mevc, and the tracks must have a track fit  $\chi^{2} < 3$ and meet at a common vertex ($\chi^{2}/\mathrm{N_{dof}} < 15$). The dedicated trigger algorithm was not enabled for the full 2011 period, resulting in an integrated luminosity of $0.65\mathrm{pb^{-1}}$ in this analysis. The remainder of the \Lc\ selection is performed at the software level, and imposes a \Lc\ mass window of $2185 < m([pK^{-}\pi^{-}]_{\Lambda_{c}^{+}}) < 2385$ \mevcc while placing a number of kinematic cuts on the candidates and particle identification (PID) requirements on the daughter tracks.

The \Xicc\ candidates are then reconstructed by pairing the reconstructed \Lc\ with two tracks which have been identified as a $K^{-}$ and $\pi^{+}$. The particles are required to point to a common vertex which is displaced from the PV. The kaon and pion tracks should also not have originated from the direction of the primary vertex and are required to have $p_{T} < 250$ \mevc. A further multivariate selection is then applied to these candidates to improve the purity of the sample. An artificial neural network is implemented utilising the TMVA package \cite{Therhaag:2010zz}. The input variables are chosen as to display minimum \Xicc\ lifetime dependence. The network is trained on simulated \Xicc\ signal samples and on $\delta m$ sideband data which is within 200 \mevcc\ of the $\delta m$ signal window.

The full selection has a limited efficiency for low \Xicc\ lifetime hypotheses. This is primarily attributable to the requirements that the reconstructed \Xicc\ vertex must be displaced from the primary vertex, and that the impact parameters of the kaon and pion should be significant with respect to the primary vertex. This analysis is therefore insensitive to $\Xi_{c}$ resonances which decay strongly to the same final state.

\section{Yield extractions}
To extract $N_{\mathrm{norm}}$ an extended maximum likelihood fit is performed to the $pK^{-}\pi^{+}$ mass spectrum. The signal shape is parameterised as the sum of two Gaussian functions with a shared mean and the background is parameterised as a first-order polynomial. The selected \Lc\ yield in the full analysis is $N_{\mathrm{norm}} = (818 \pm 7) \times 10^3$, with a mass resolution of $\approx6$ \mevcc.

The \Xicc\ yield is extracted from the $\delta m$ distribution for a number of \dm\ hypotheses. The method requires sufficient knowledge of the signal mass resolution to define a signal window, but beyond that requires no further information on the \Xicc\ lineshape. This is determined with a fit to the simulated signal, parameterising the signal as the sum of two Gaussian functions with a shared mean. The resolution is determined to be $\approx 4$\mevcc. For each investigated $\delta m$ a narrow signal region is defined as $2273 < m([pK^{-}\pi^{-}]_{\Lambda_{c}^{+}}) < 2303 \mathrm{MeV/c^{2}}$ and $| \delta m - \delta m_{0}| < 10\mathrm{MeV/c^{2}}$. Candidates outside this window are used to estimate the expected background within the signal window, and this is subtracted from the number of candidates inside the window to calculate the signal yield for that value of $\delta m$.

Two methods following this procedure are used. The first is an analytic two dimensional sideband subtraction, which uses a $5 \times 5$ array of non-overlapping, variable size tiles centred on the signal region with total width of $80\mevcc$ in $m([pK^{-}\pi^{-}]_{\Lambda_{c}^{+}})$ and total width $200\mevcc$ in \dm. The combinatoric background is parameterised by a two-dimensional quadratic function while the \Lc\ component is described by the product of a signal peak in $m([pK^{-}\pi^{-}]_{\Lambda_{c}^{+}})$ and a quadratic function in \dm. The background distribution is then extracted from the $24$ non-central bins and the integral of this distribution over the signal box (central bin) is evaluated, extracting the background and associated statistical error. A second, cross check method is also employed by imposing a narrow \Lc\ mass window on all candidates and reducing the problem to a one-dimensional \dm\ distribution.

\section{Efficiency corrections and systematics}
The efficiency ratios in the analysis are calculated using a variety of data-driven methods and methods utilising simulated data. The kinematic distributions of \Xicc\ at the LHC are unknown. The simulation used in this analysis is generated according to the \texttt{GENXICC} \cite{Chang:2009va} model, and with $m(\Xi_{cc}^{+}=3500~\mevcc)$ and $\tau_{\Xi_{cc}^{+}}= 333~\mathrm{fs}$. The efficiency ratio may be factorised into the following components:
\begin{equation}
\frac{\epsilon_{\mathrm{norm}}}{\epsilon_{\mathrm{sig}}} = 
\frac{\epsilon_{\mathrm{norm}}^{\mathrm{acc}}}{\epsilon_{\mathrm{sig}}^{\mathrm{acc}}}
\frac{\epsilon_{\mathrm{norm}}^{\mathrm{sel|acc}}}{\epsilon_{\mathrm{sig}}^{\mathrm{sel|acc}}}
\frac{\epsilon_{\mathrm{norm}}^{\mathrm{PID|sel}}}{\epsilon_{\mathrm{sig}}^{\mathrm{PID|sel}}}
\frac{1}{\epsilon_{\mathrm{sig}}^{\mathrm{ANN|PID}}}
\frac{\epsilon_{\mathrm{norm}}^{\mathrm{trig|PID}}}{\epsilon_{\mathrm{sig}}^{\mathrm{trig|ANN}}}
\end{equation}
where the efficiencies correspond to the acceptance (acc), the reconstruction and selection excluding the PID and ANN requirements (sel), the particle identification requirements (PID), the ANN selection for the signal mode only (ANN), and the trigger (trig). Most of these are evaluated with the use of simulated \Xicc\ and \Lc\ decays. Due to known discrepancies between the data and simulation corrections to these efficiencies are required. The efficiency of the PID requirements, the tracking and the calorimeter hardware trigger are evaluated with the use of data-driven calibration techniques.


As the \Xicc\ mass and lifetime are \emph{a priori} unknown, it is necessary to re-weight the simulated events to evaluate the efficiencies for a variety of potential \Xicc\ properties. In the case of the \Xicc\ lifetime, the simulated events are re-weighted with a different exponential distribution and the efficiency is recalculated. In the case of the \Xicc\ mass, simulated data is generated under two other mass hypotheses, $m(\Xi_{cc}^{+}=3300\mev)$ and $m(\Xi_{cc}^{+}=3700\mev)$ without simulating interactions with the detector. The kinematics of the \Xicc\ daughters in the primary simulated data are re-weighted to match the distributions of the low and high mass simulation data and the efficiency is redetermined. Defining the event sensitivity $\alpha$ as 
\begin{equation}
\alpha \equiv \frac{\epsilon_{\mathrm{norm}}}{N_{\mathrm{norm}}\epsilon_{\mathrm{sig}}}
\end{equation}
such that $ R = \alpha N_{\mathrm{sig}}$, it was found that $\alpha$ varies strongly with \Xicc\ lifetime and weakly with \Xicc\ mass. 
%

The dominant uncertainty in the analysis is the statistical uncertainty on the measured signal yield, and systematic uncertainties on $\alpha$ have limited effects on the expected upper limits. The dominant systematic uncertainty in the analysis is due to the limited sample size of simulated events used in the efficiency corrections. Smaller systematic effects are also associated with the data-driven efficiency calibration methods. The systematic uncertainty depends on the \Xicc\ lifetime and mass hypotheses used. Adding these effects in quadrature an overall systematic for the analysis of $26\%$ is assigned.

\section{Results and conclusions}
Tests for \Xicc\ signals are carried out at 1\mevcc steps across the full \dm\ range. For each value, yields for signal and background are extracted as in Sec. $4$. Local significances are then calculated as
\begin{equation}
\mathcal{S} (\delta m) \equiv \frac{N_{S+B}-N_{B}}{\sqrt{\sigma^{2}_{S+B} + \sigma^{2}_{B}}}
\end{equation}
where $\sigma^{2}_{S+B}$ and $\sigma^{2}_{B}$ are the statistical uncertainties on the signal yield and the expected background. The look elsewhere effect \cite{Lyons:1900zz} is taken into account to correct for a global significance. A large number of simulated background-only pseudo-experiments are generated and the full analysis procedure is applied to each. The global $p$-value for a given $S$ is then the fraction of the total simulated experiments which contained an equal or larger local significance at any value of \dm. If no signal excess corresponding to a global significance of $3\sigma$ is observed, upper limits on $R$ are quoted using the $CL_{S}$ method \cite{Read:2002hq}. 

\begin{figure}[htb]
\centering
\includegraphics[width=0.72\textwidth]{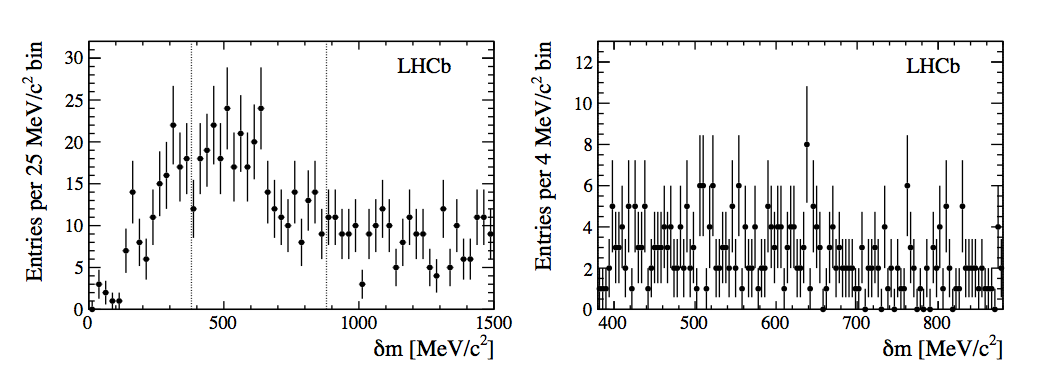}
\caption{The \dm\ distribution requiring $2273 < m([pK^{-}\pi^{-}]_{\Lambda_{c}^{+}}) < 2303 \mev$. The right plot shows the highlighted range in the left with a finer binning.}
\label{fig:dm}
\end{figure}

\begin{figure}[H]
\centering
\includegraphics[width=0.52\textwidth]{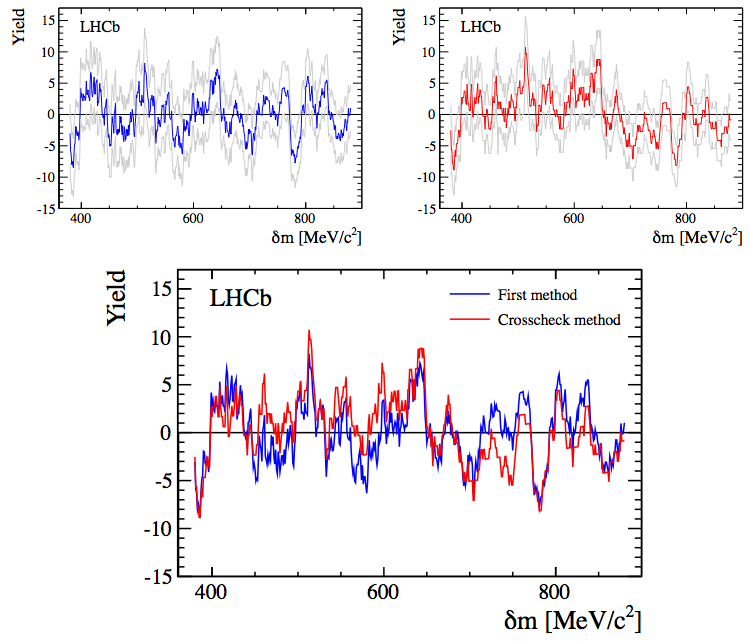}
\caption{The measured signal yields in \dm. The upper plots show the yields for the primary extraction method (left) and the cross-check method (grey lines are $\pm1~\sigma$ statistical error bands). Lower plot shows both methods plotted together, indicating good agreement.}
\label{fig:yields}
\end{figure}

\begin{figure}[H]
\centering
\includegraphics[width=0.52\textwidth]{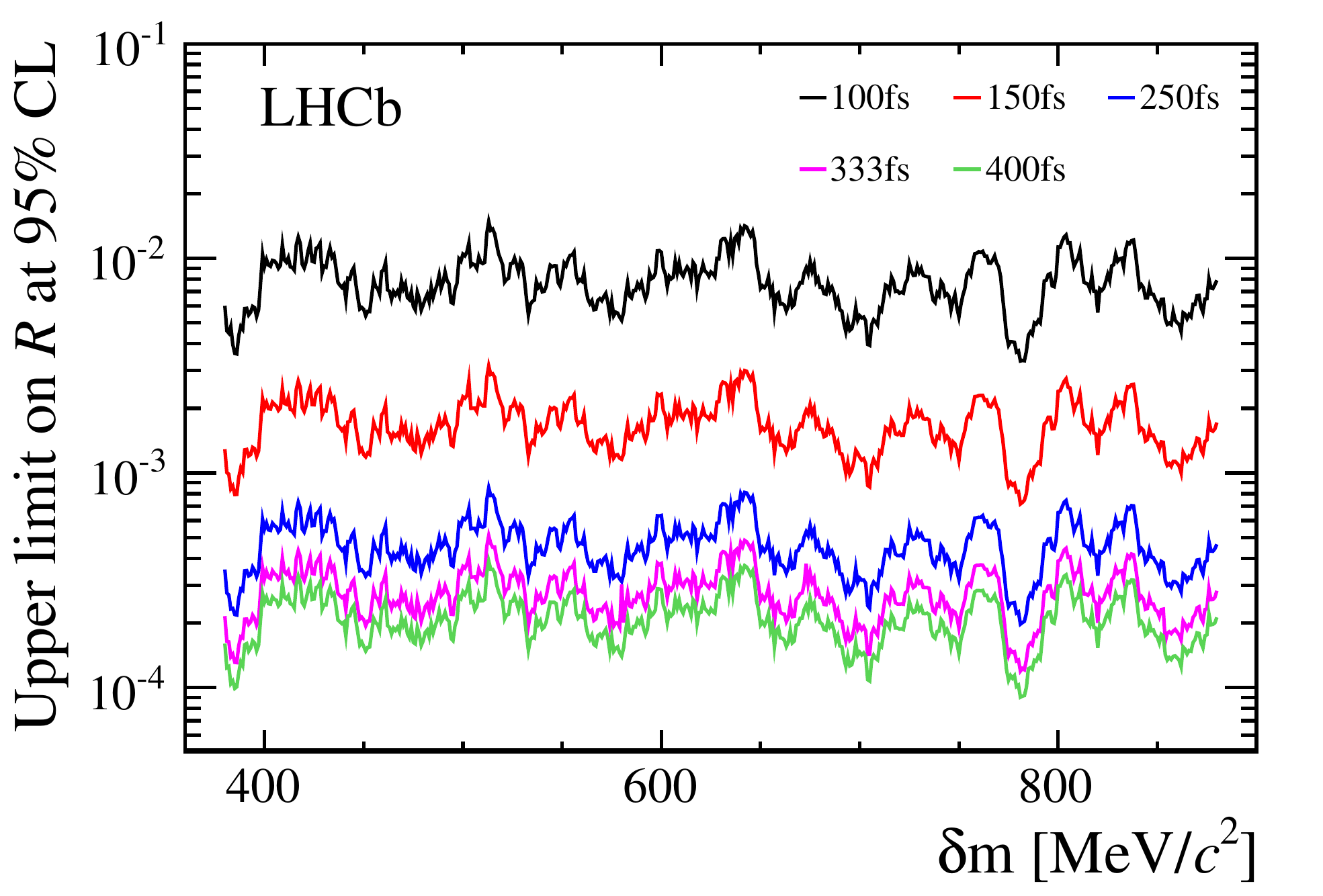}
\caption{Upper limits on $R$ for a number of \Xicc\ lifetime hypotheses.}
\label{fig:uls}
\end{figure}

The \dm\ distribution is shown in Fig. \ref{fig:dm}, and the estimated signal yield in Fig. \ref{fig:yields}. The largest local significance observed is at $\delta m = 513\mev$ corresponding to a local significance $\mathcal{S}=1.5~\sigma$ ($2.2~\sigma$ in the 1D cross-check fit). This corresponds to a global $p$-value of $99~\%$ (53~\%). It is therefore concluded that no significant excess is observed. Upper limits on $R$ are given in Fig. \ref{fig:uls} across the \dm\ distribution for a variety of lifetime hypotheses.

\clearpage
\Acknowledgements
We express our gratitude to our colleagues in the CERN
accelerator departments for the excellent performance of the LHC. We
thank the technical and administrative staff at the LHCb
institutes. We acknowledge support from CERN and from the national
agencies: CAPES, CNPq, FAPERJ and FINEP (Brazil); NSFC (China);
CNRS/IN2P3 and Region Auvergne (France); BMBF, DFG, HGF and MPG
(Germany); SFI (Ireland); INFN (Italy); FOM and NWO (The Netherlands);
SCSR (Poland); MEN/IFA (Romania); MinES, Rosatom, RFBR and NRC
``Kurchatov Institute'' (Russia); MinECo, XuntaGal and GENCAT (Spain);
SNSF and SER (Switzerland); NAS Ukraine (Ukraine); STFC (United
Kingdom); NSF (USA). We also acknowledge the support received from the
ERC under FP7. The Tier1 computing centres are supported by IN2P3
(France), KIT and BMBF (Germany), INFN (Italy), NWO and SURF (The
Netherlands), PIC (Spain), GridPP (United Kingdom). We are thankful
for the computing resources put at our disposal by
Yandex LLC (Russia), as well as to the communities behind the multiple open
source software packages that we depend on.

\end{document}